\documentclass[aps,pra,amsfonts,twocolumn,nofootinbib,groupedaddress]{revtex4}

\usepackage{color}
\usepackage{graphics,epsfig}

\newcommand {\be} {\begin{equation}}
\newcommand {\ba} {\begin{eqnarray}}
\newcommand {\ee} {\end{equation}}
\newcommand {\ea} {\end{eqnarray}}

\begin{document}

\title{Proton structure corrections to electronic and muonic hydrogen hyperfine splitting}

\author{Carl E.\ Carlson$^{(a)}$, Vahagn Nazaryan$^{(b)}$, and Keith Griffioen$^{(a)}$}
\affiliation{
$^{(a)}$Department of Physics, College of William and Mary, Williamsburg, VA 23187, USA\\
$^{(b)}$Center for Advanced Medical Instrumentation,
Department of Physics, Hampton University, Hampton,
VA 23668}

\date{May 15, 2008}

\begin{abstract}

We present a precise determination of the polarizability and other proton structure dependent contributions to the hydrogen hyperfine splitting, based heavily on the most recent published data on proton spin dependent structure functions from the EG1 experiment at the Jefferson Laboratory.  As a result, the total calculated hyperfine splitting now has a standard deviation slightly under 1 part-per-million, and is about 1 standard deviation away from the measured value.  We also present results for muonic hydrogen hyperfine splitting, taking care to ensure the compatibility of the recoil and polarizability terms. 

\end{abstract}

\maketitle


\section{Introduction}


In this article, we consider precision calculation of the hyperfine splitting (hfs) of hydrogen, with the goals of calculating the hfs to a part-per-million (ppm) accuracy for ordinary (electronic) hydrogen and of extending the calculation to the muonic hydrogen case.  

Experimentally, the hfs of the hydrogen ground state is known to 13 significant figures in frequency units~\cite{Karshenboim:1997zu},
    \be
    E_{\rm hfs}(e^-p) = {\rm 1\ 420.405\ 751\ 766\ 7  (9)\ MHz} \,.
    \ee
On the theoretical side, at the level of a ppm accuracy, the QED corrections are not in question.  Rather, achieving the stated accuracy requires better evaluation of corrections from the finite size of the proton.  Finite size corrections come from two-photon exchange, Fig.~\ref{fig:one}, where there is the possibility that the photons are individually hard and can see deeply into the proton.  For one photon exchange, proton structure plays no role at the ppm level because the momentum transfer is necessarily very low.

Presently, our ability to numerically deal with quantum chromodynamics (QCD), the theory that governs how matter is bound together to form a proton, is insufficient to calculate proton structure corrections to the desired accuracy.  Instead,  the corrections can be related to proton structure information measured in electron-proton scattering.  The information is codified in terms of Pauli and Dirac form factors $F_1(Q^2)$ and $F_2(Q^2)$ for the elastic case and structure functions $g_1(\nu,Q^2)$ and $g_2(\nu,Q^2)$ for the spin-dependent inelastic case.  Here $Q^2 = - q^2$, where $q$ is the 4-momentum transferred from the electron and $\nu$ is the energy transferred from the electron in the lab frame; one can also use $x=Q^2/(2m_p\nu)$, where $m_p$ is the proton mass. 

Recently reported~\cite{Prok:2008ev} data on $g_1(\nu,Q^2)$ from JLab spur the present study.   The data are from the EG1 collaboration, and extend the measurements of $g_1(\nu,Q^2)$ down to much lower $Q^2$ than previously available.  The relations between the hfs and $g_1(\nu,Q^2)$ weight heavily on the low $Q^2$ data, so the latest data, which include a careful analsys of statistical and systematic errors, lead to a more accurate and reliable hfs calculation.

The proton structure dependent corrections can be divided into Zemach, recoil, and polarizability corrections, to be defined shortly.  The first two depend entirely on elastic intermediate states in the two-photon exchange, and all contributions from inelastic intermediate states are in the third.  Our main, though not exclusive, focus will be upon the polarizability corrections, which contain the dependence upon $g_1(\nu,Q^2)$ and which have had larger statistical and systematic uncertainties limits than the other two terms.  The situation has now improved and we will find that the uncertainty in the polarizability corrections is now comparable to the uncertainty in the elastic structure dependent terms.  Hence we shall evaluate these also, using up-to-date form factors, and discuss the uncertainty limits in the calculations of all terms.  

For muonic hydrogen hfs, there is currently no measurement, but one may be possible~\cite{Antognini:2005fe}, so it is appropriate to quote a calculated result.   In the muon case, the structure dependent corrections are in total larger than the QED corrections, because the former have a lepton mass proportionality, while the latter are to a first approximation independent of the lepton mass.

The muon case prompts a discussion of the definitions of the recoil and polarizability corrections. The  sum of all proton structure corrections is unambiguous.  However, the separation between the recoil and polarizability corrections depends upon a protocol.  The issue is that the elastic and inelastic corrections separately have (after an overall lepton mass, $m_\ell$, is factored out) logarithmic divergences in the $m_\ell \to 0$ limit.  For convenience, the polarizability corrections have been defined~\cite{Drell:1966kk} by taking the inelastic corrections and adding an elastic-looking term to cancel the logarithmic singularity.  An identical term is subtracted from the recoil corrections, and the overall sum is unchanged.   

The term added to form the polarizability correction must satisfy the criteria that it cancel the existing  $m_\ell\to 0$ divergence, and that it introduce no new divergence.  This does not uniquely fix the residual non-divergent part of the term.  For the electron case,  the choice is standard.  Hence, one can in principle add calculations of electronic hydrogen polarizability and recoil corrections from different sources without worry.  For the massive lepton case, it appears that there are two different protocols, which agree in the $m_\ell\to 0$ limit but not otherwise.  Hence, there is a need for care in combining muonic hydrogen calculations from different sources, or else for a unified calculation of all the proton structure dependent terms, as we do here.

Our calculations and results are detailed in Section~\ref{sec:calc}.  The relevant formulas are first summarized and discussed, followed by numerical evaluations for the electronic and muonic hydrogen systems.  Section~\ref{sec:end} summarizes our conclusions.


\section{Formulas and calculations}				\label{sec:calc}



\subsection{Formulas and calculations}


The calculated hyperfine splitting can be given as~\cite{Volotka:2004zu,dupays}
\ba
    E_{\rm hfs}(\ell^-p) &=&
    \big (1+\Delta_{\rm QED}+\Delta_{\rm hvp}^p
        \nonumber \\[1ex]
    && +\ \Delta_{\mu{\rm vp}}^p+\Delta_{\rm weak}^p+\Delta_{\rm S} \big)
    \, E_F^p \,,
\ea
where lepton $\ell^-$ is either $e^-$ or $\mu^-$ and the Fermi energy is
\be
    E_F^p=\frac{8 \alpha^3 m_r^3 }{3\pi} \mu_B\mu_p
    	=	\frac{16  \alpha^2}{3} \frac{\mu_p}{\mu_B}  
		\frac{ R_\infty }{ \left( 1 + m_\ell/m_p \right)^3 } 	\,.	
\ee
Mass $m_r = m_\ell m_p / (m_p + m_\ell)$ is the reduced mass and $R_\infty$ is the Rydberg constant (in frequency units).  By convention, the Bohr magneton $\mu_B$ is inserted for the lepton and the measured magnetic moment $\mu_p$ is used for the proton.   The constants on the right-hand-side are well enough known to evaluate the Fermi energy to $0.01$ ppm.

The first four corrections are due to QED, hadronic vacuum polarization, muonic vacuum polarization, and weak interactions ($Z^0$ exchange).  They are well enough known not to require discussion here.

The proton structure dependent corrections are 
\be
\Delta_{\rm S}  = \Delta_Z + \Delta_R^p + \Delta_{\rm pol} \,.
\ee
The subscripts stand for ``Zemach,'' ``recoil,'' and ``polarizability.''   The measured value of $E_{\rm hfs}(e^-p)$ and calculated values of other quantities implies a ``target value'' $\Delta_S = -32.77 \pm 0.01$ ppm for ordinary hydrogen~\cite{Volotka:2004zu,dupays}.

\begin{figure}

\centerline{ \includegraphics[height=20mm] {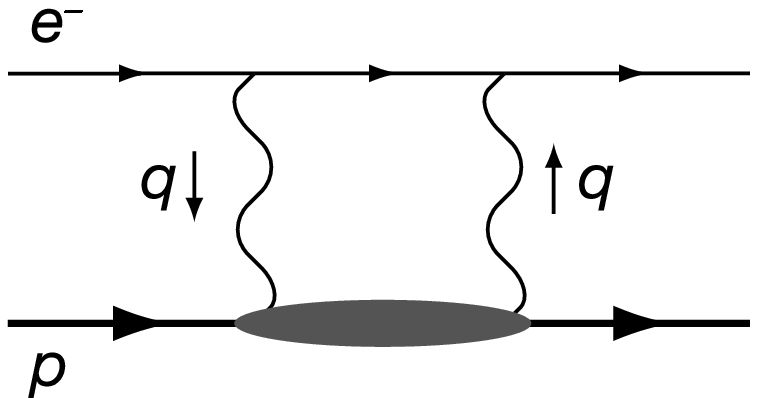} }
\vskip 5mm
\centerline{ \includegraphics[width=3.3in] {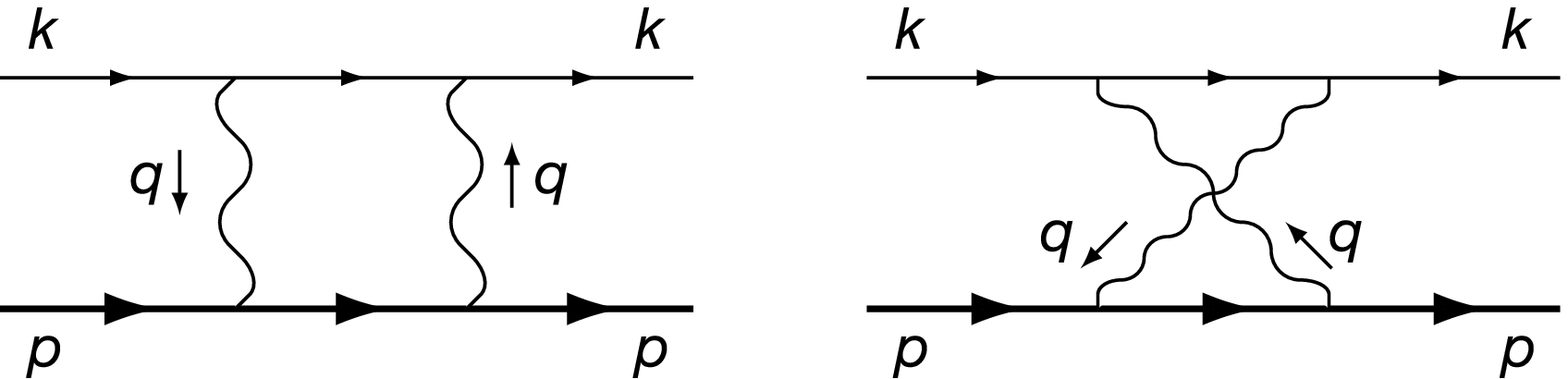} }

\caption{Upper part: the full box;  lower part: the box with elastic intermediate states only.}

\label{fig:one}
\end{figure}

The structure dependent corrections can all obtained by a dispersive calculation of the two-photon exchange diagram (Fig.~\ref{fig:one}), as pioneered by Iddings~\cite{Iddings} and by Drell and Sullivan~\cite{Drell:1966kk}.  The reason for separating the result into three terms is partly to shorten individual formulas and partly for historical reasons.  We only quote the results, reserving a discussion of the derivation, particularly for the massive lepton case, for a later report.

We start with the polarizability corrections.  They are usually given in the limit $m_\ell \to 0$~\cite{IddingsP,Iddings,Drell:1966kk,DeRafael:mc,Gnaedig:qt}.  To our knowledge, the only previous exception is in the article of Cherednikova, Faustov, and Martynenko~\cite{Faustov:2001pn}.  Including the lepton mass, our result for the polarizability corrections is
\ba
\Delta_{\rm pol} 
	&=& \frac{ \alpha m_\ell } { 2(1+\kappa_p) \pi m_p } \left( \Delta_1 + \Delta_2 \right) ,
\ea
with 
\ba
\label{eq:DeltaMassive}
\Delta_1 &=& 
	\int_0^\infty	 \frac{ d Q^2 }{ Q^2 } \Bigg\{
	\beta_1(\tau_\ell) F_2^2(Q^2)
				\\
	&& \hskip 4 em +\ \frac{ 8 m_p^2}{Q^2}  \int_0^{x_{th}}  dx \ 
	\tilde\beta_1(\tau,\tau_\ell)
	g_1(x,Q^2)	\Bigg\}
				\nonumber	\\[1.25ex]
\Delta_2 &=&  -24 m_p^2 
	\int_0^\infty	 \frac{ d Q^2 }{ Q^4 }
	\int_0^{x_{th}}  dx	\ 
	\tilde\beta_2(\tau,\tau_\ell) \ 
	g_2(x,Q^2)				\,,		\nonumber 
\ea
where $\kappa_p$ is the proton anomalous magnetic moment in nuclear magnetons, $x_{th} = Q^2 / (2 m_p m_\pi + m_\pi^2 + Q^2)$ with $m_\pi$ the charged pion mass, and
\ba
\tilde\beta_1(\tau,\tau_\ell) &=& 
	\frac{x^2 \beta_1(\tau) - (m_\ell^2 / m_p^2) \beta_1(\tau_\ell)}{x^2 - m_\ell^2/ m_p^2}
		\nonumber \\
\tilde\beta_2(\tau,\tau_\ell) &=&
	\frac{x^2 \left[ \beta_2(\tau) - \beta_2(\tau_\ell) \right] }{x^2 - m_\ell^2/ m_p^2}  \,.
\ea
The plain $\beta_{1,2}$ auxiliary functions, introduced by De Rafael~\cite{DeRafael:mc}, are 
\begin{eqnarray}
\beta_1(\tau) &=& -3\tau + 2\tau^2
    + 2(2-\tau)\sqrt{\tau(\tau+1)} 	\,,
                    \nonumber \\
\beta_2(\tau) &=& 1 + 2\tau - 2 \sqrt{\tau(\tau+1)}		\,,
\end{eqnarray}
which have limits
\begin{eqnarray}
\beta_1(\tau) &=& \left\{
        \begin{array} {cl}
        4 \sqrt{\tau} + {\cal O}(\tau)	 \,, & \ \tau \to 0  \\[1.25ex]
        \frac{9}{4} \left( 1 - \frac{5}{18} \frac{1}{\tau} + \frac{7}{48} \frac{1}{\tau^2} + \ldots \right)\,,
            &      \  \tau\to\infty
        \end{array}
        \right.		\,,
                    \nonumber 	\\[1.5ex]
\beta_2(\tau) &=& \left\{
        \begin{array} {cl}
        1 + {\cal O}(\sqrt{\tau})\,, & \ \tau \to 0  \\[1.25ex]
        0 + \frac{1}{4} \frac{1}{\tau}  - \frac{1}{8} \frac{1}{\tau^2} + \ldots \,,
            &      \  \tau\to\infty
        \end{array}
        \right.					\,.
\end{eqnarray}

\noindent and are used with the notations
\be
\tau      \equiv \frac{\nu^2}{Q^2}	 \,,	\quad
\tau_\ell \equiv \frac{Q^2}{4m_\ell^2} \,,	\quad
\tau_p \equiv \frac{Q^2}{4m_p^2}	\,.
\ee
Finally, $F_2(Q^2)$ is the (elastic) Pauli form factor, normalized by $F_2(0)= \kappa_p$.

The polarizability terms come mainly from inelastic intermediate states in Fig.~\ref{fig:one}; hence the appearance of the structure functions $g_{1,2}$.  The term containing $F_2$ is the term described in the introduction, which is inserted to cancel the divergence that appears in the $g_1$ term of $\Delta_1$ in the massless lepton limit. As $m_\ell\to 0$, one can show that $\gamma_{1,2} \to \beta_{1,2}$ and $\beta_1(\tau_\ell)\to 1$, and further, for $Q^2 \to 0$, $\beta_1(\tau) \to 1$. Then the Gerasimov-Drell-Hearn~\cite{Gerasimov:1965et,Drell:1966jv} sum rule,
\be
\lim_{Q^2\to 0}
    \frac{8m_p^2}{Q^2} \int_0^{x{\rm th}} dx \, g_1(x,Q^2)
        = -\kappa^2_p
\ee
both ensures that the second term of $\Delta_1$ diverges in the massless limit, and that the first term will regularize it.  

Our polarizability correction agrees with~\cite{Faustov:2001pn} for the $g_{1,2}$ terms, which are unique.  Some choice is possible for the $F_2^2$ terms, and we and~\cite{Faustov:2001pn}  made different choices.  Further explanation of this point joins the discussion of the recoil correction, below.

The Zemach corrections are~\cite{Zemach}
\ba
\Delta_Z =  -2 \alpha m_r r_Z  \left( 1+\delta_Z^{rad} \right)	,
\ea
where $r_Z$ is the Zemach radius
\be
r_Z = -  \frac{4}{\pi} 
	\int_0^\infty \frac{dQ}{Q^2} \left( G_E(Q^2) \frac{G_M(Q^2)}{1+\kappa_p} -1 \right)  .
\ee
The electric and magnetic Sachs form factors are
\ba
G_M(Q^2) &=& F_1(Q^2) + F_2(Q^2)	,		\nonumber  \\
G_E(Q^2) &=& F_1(Q^2) -  \frac{Q^2}{4m_p^2} F_2(Q^2)   ,
\ea
and the Dirac form factor is normalized with $F_1(0) = 1$.

The extra radiative correction $\delta_Z^{\rm rad}$ is given in~\cite{Bodwin:1987mj,Karshenboim:1996ew}. For the dipole form factor,  $G_{E,M}(Q^2) \propto (1+Q^2/\Lambda^2)^{-2}$, 
one finds
 $\delta_Z^{\rm rad} = (\alpha/3\pi)
    \left[ 2 \ln ( \Lambda^2/ m_e^2 ) - 4111/420 \right] = 0.0153$, using the standard value
$\Lambda^2 = 0.71$ GeV$^2$.   For other form factors that we use, the changes in $\delta_Z^{\rm rad}$ have a 0.01 ppm or smaller effect upon the hyperfine splitting.

The leading order recoil corrections are~\cite{Bodwin:1987mj,Martynenko:2004bt}
\begin{widetext}
 \ba
 \label{eq:recoil}
 \Delta_R^p &=&  \frac{2 \alpha m_r}{\pi m_p^2} 
	\int_0^\infty dQ \,
	F_2(Q^2) \frac{G_M(Q^2)}{1+\kappa_p}
					\nonumber	\\[1.25ex]
&& + \ \  \frac{ \alpha m_\ell m_p} { 2(1+\kappa_p) \pi (m_p^2-m_\ell^2) }
	\Bigg\{
	\int_0^\infty    \frac{ d Q^2 }{ Q^2 }
	\left(     \frac{\beta_1(\tau_p)-4\sqrt{\tau_p}}{\tau_p} 
		- \frac{\beta_1(\tau_\ell)-4\sqrt{\tau_\ell}}{\tau_\ell}		\right)
	F_1(Q^2) G_M(Q^2)  
					\nonumber	\\[1.25ex]
	&&	\hskip 3.8cm
	+ \ 3 \int_0^\infty  \frac{ d Q^2 }{ Q^2 } \   \Big( \beta_2(\tau_p) - \beta_2(\tau_\ell)	\Big) \ 
	F_2(Q^2) G_M(Q^2)	\Bigg\}
					\nonumber	\\[1.25ex]
&& - \ \ \frac{ \alpha m_\ell } { 2(1+\kappa_p) \pi m_p }
	\int_0^\infty	 \frac{ d Q^2 }{ Q^2 } \ \beta_1(\tau_\ell) F_2^2(Q^2)		\,.
 \ea
 
\end{widetext}

Factoring out an overall $m_\ell$, there remain recoil terms that diverge like $\ln(m_\ell)$ as $m_\ell \to 0$; hence the $m_\ell \to 0$ limit is not taken.  Further, and in contrast to the Zemach  corrections, the recoil corrections are not zero in the static and pointlike proton limits.  [The static limit neglects the $Q^2$ dependence of the form factors, so that $F_1(Q^2) \to F_1(0) = 1$ and $F_2(Q^2) \to F_2(0) = \kappa_p$;  the pointlike limit additionally takes $\kappa_p \to 0$.]  Thus, part of the recoil correction is structure independent.  However, that they are overall structure dependent is clear, and so it is proper to include them here along with $\Delta_Z$ and $\Delta_{\rm pol}$.

Notice that the last term in the recoil correction is the negative of the $F_2$ term from the $\Delta_1$ polarizability correction.  These are the terms that were added and subtracted to ensure that  $\Delta_1$ contained no divergence in the massless lepton limit.

We specify the term here using a historical criterion.  An alternative non-dispersive calculation of the elastic contributions alone, the lower part of Fig.~\ref{fig:one},  inserts
photon-proton-proton vertices
\be
\label{eq:vertex}
\Gamma_\nu = \gamma_\nu F_1(Q^2) + \frac{i}{2m_p} \sigma_{\nu\rho}q^\rho F_2(Q^2) \,,
\label{eq:Gamma}
\ee

\noindent  for incoming $q$, and does the loop integral directly.  In modern times, one should hesitate to do the calculation this way because there is no reason to think the vertex representation is correct when the intermediate proton is off shell.     (The dispersive calculation is not subject to the same criticism because it obtains the real part of the two-photon corrections from the imaginary part, which only requires knowing the vertices when all protons are on-shell.)  However, the direct calculation is historically older than the dispersive one, and is still often quoted; for relatively modern source see~\cite{Martynenko:2004bt}.   It is possible to choose the $F_2^2$ terms in the polarizability and recoil corrections to cancel the zero mass divergence in one case and give the historical result in the other, and that is the choice we made.   Ref.~\cite{Faustov:2001pn}, which showed only the polarizability term, differs from us in the $F_2^2$ terms in $\Delta_{\rm pol}$ and this can be traced to a different choice early on~\cite{Zinovev1970}.


\subsection{Ordinary hydrogen polarizability corrections}


For electronic hydrogen, take the $m_\ell \to 0$ limit to obtain the well-known result~\cite{IddingsP,Iddings,Drell:1966kk,DeRafael:mc,Gnaedig:qt},
\ba
  \Delta_1 &=& \frac{9}{4}\int_0^\infty \frac{dQ^2}{Q^2}\left\{F_2^2(Q^2) + \frac{8m_p^2}{Q^2}
B_1(Q^2)\right\}   \,,
										\nonumber\\
\Delta_2 &=& -24m_p^2\int_0^\infty \frac{dQ^2}{Q^4}B_2(Q^2) .
	\label{eq:Delta}
\ea

\noindent with
\ba
B_1(Q^2) &=& \frac{4}{9} \int_0^{x_{\rm th}} dx \, \beta_1(\tau) g_1(x,Q^2)  \,,  \nonumber \\
B_2(Q^2) &=& \int_0^{x_{\rm th}} dx \, \beta_2(\tau) g_2(x,Q^2)  \,.
\label{eq:B}
\ea

Information on $g_1(x,Q^2)$ is obtained from polarized electron on polarized proton inelastic scattering, and the lowest $Q^2$ data come from the EG1 experiment at JLab, reported in~\cite{Prok:2008ev,Dharmawardane:2006zd} with data details posted on the High Energy Physics database at Durham University (UK).  The EG1 data have beam energies of $1.6$ and $5.7$ GeV, and give $g_1(\nu,Q^2)/F_1(\nu,Q^2)$ [$F_{1,2}(\nu,Q^2)$ are the spin-independent structure functions] for 28 $Q^2$ bins with central values from $0.0496$ to $4.96$ GeV$^2$, and with $W$ ranging from threshold to about $1.65$ GeV for the lower $E_{\rm beam}$ data and from threshold to about $3.1$ GeV for the higher $E_{\rm beam}$ data.  We obtain $F_1(\nu,Q^2)$ in the resonance region from the Christy and Bosted parameterization~\cite{Christy:2007ve}, and in the scaling region from combining the $F_2(\nu,Q^2)$ fit of the NMC collaboration~\cite{Arneodo:1995cq} with the fit to $R$ (the ratio of longitudinal and transverse photon cross sections) from SLAC E143~\cite{Abe:1998ym}.  Where there is no EG1 data, we use fits to $g_1$ from Simula {\it et al.}~\cite{Simula:2001iy} in the resonance region and from SLAC E155~\cite{Anthony:2000fn} in the scaling region.

For $Q^2$ below $Q_1^2 = 0.0452$ GeV$^2$ (the lower edge of the lowest bin) there is no data, and we complete the integrals by interpolating data between higher $Q^2$ and zero $Q^2$.  For  $B_1$, which is proportional to $Q^2$ as $Q^2 \to 0$, this is possible because the GDH sum rule~\cite{Gerasimov:1965et,Drell:1966jv} fixes the slope,  so that for small $Q^2$,
\be
\label{eq:b1expansion}
B_1(Q^2) = -\frac{\kappa_p^2}{8m_p^2} Q^2 + c_{1B} Q^4 + \ldots,
\ee
We obtain $c_{1B} = 4.94 \pm 0.30 \pm 1.22$ GeV$^{-4}$ by fitting to $B_1(Q^2)$ in the data region below $Q^2 = 0.3$ GeV$^2$.   (This is somewhat larger than the $c_1$ we quote in~\cite{Nazaryan:2005zc}, partly because we are here fitting $B_1$ instead of $\Gamma_1$, but more because of improvements in the data.)  The contribution to $\Delta_1$ from the low $Q^2$ range is thus
\be
\Delta_1[0,Q_1^2] = \left\{ -\frac{3}{4} \kappa_p^2 r_P^2 + 18 m_p^2 c_{1B} \right\} Q_1^2
	+ {\cal O}(Q_1^4 r_p^4),
\ee
where $r_P$ is the Pauli radius of the proton, coming from the expansion of $F_2(Q^2)$.

The structure function $g_2$ gives a small contribution to the hfs, because the auxiliary function $\beta_2$ is small for the kinematics where the $g_2$ integral has its main support.  This is fortunate, since $g_2$ for the proton is not well measured.  The $\Delta_2$'s in our Tables are based on a model for $g_2$ provided by the EG1 collaboration, which we also used in~\cite{Nazaryan:2005zc} and which is heavily based on the MAID parameterization~\cite{Drechsel:2007if} of existing photo- and electroproduction data.  Given the overall lack of data that is specific to $g_2$, we assigned $100$\% error limits to the $\Delta_2$ determnation; even so, the contribution to the overall uncertainty of $\Delta_{\rm pol}$ is not large.  A more detailed discussin of $\Delta_2$ is given in Sec.~\ref{subsec:deltatwo}.

\begin{table}[t]
\caption{Contributions to $\Delta_{\rm pol}$ for electron case.  Statistical, systematic, and modeling errors, in that order, are given in the parentheses and discussed in the text.   }
\begin{ruledtabular}
\begin{tabular}{llcc}
Term		& $Q^2$ (GeV$^2$)	&  From  &  Value w/AMT~\cite{Arrington:2007ux} $F_2$	\\
\hline
$\Delta_1$	& [0, 0.0452]	&	$F_2$ \& $g_1$	& $\ \  1.35 (0.22) (0.87) \ \ (\ )\ \  $	\\
			& [0.0452, 20]\ \ \ &	$F_2$			& $\ \  7.54 \ \ (\ )\ \  (0.23) \ \ (\ )\ \  $	\\
			& 			&	$g_1$	& $-0.14 (0.21) (1.78) (0.68)$\\
			& [20, $\infty$]	&	$F_2$	& $\ \  0.00  \ \ (\ )\ \  (0.00) \ \ (\ )\ \ $	\\
			&			&	$g_1$	& $\ \  0.11  \ \ (\ )\ \  \ \ (\ )\ \  (0.01)$  \\[1.25ex]
total $\Delta_1$&			&			& $\ \  8.85 (0.30) (2.67) (0.70)$	\\[1.5ex] 
$\Delta_2$	& [0, 0.0452]	&	$g_2$	& $ - 0.22  \ \ (\ )\ \  \ \ (\ )\ \  (0.22)$		\\
			& [0.0452, 20]	&	$g_2$	& $ - 0.35  \ \ (\ )\ \  \ \ (\ )\ \  (0.35)$		\\
			& [20, $\infty$]	&	$g_2$	& $ \ \ 0.00  \ \ (\ )\ \  \ \ (\ )\ \  (0.00)$	\\
total $\Delta_2$&			&			& $ - 0.57  \ \ (\ )\ \  \ \ (\ )\ \  (0.57)$	\\[1.5ex]
$\Delta_1 + \Delta_2$	&	&			& $\ \  8.28 (0.30) (2.67) (0.90)$\\[1.5ex]
$\Delta_{\rm pol}$ (ppm)		&	&		& $\ \  1.88 (0.07) (0.60) (0.20)$
\end{tabular}
\end{ruledtabular}
\label{table:pol}					
\end{table}%

Results for the polarizability correction are broken down in Table~\ref{table:pol} for one particular parameterization of the elastic form factor $F_2(Q^2)$.  In this Table, ``systematic errors'' mean systematic errors that come from the listed data~\cite{Prok:2008ev,Dharmawardane:2006zd}, and ``modeling errors'' come from error limits accompanying the parameterizations~\cite{Simula:2001iy} that we use to complete the integrals where data is lacking.  Not all of these uncertainties apply to each of the numbers in the Table, and we indicate this by leaving blanks in the parenthesis.  In the Table, the statistical errors are always combined in quadrature;  the systematic errors from $g_1$ (the error in the low $Q^2$ bin is treated as due to uncertainty in $g_1$, as it mostly is) are combined directly, and then combined in quadrature with those from the $F_2$ terms;  and the modeling errors in $\Delta_1$ and $\Delta_2$ are separately combined directly, and then combined in quadrature with each other.

Further combining the statistical, systematic, and modeling errors in quadrature gives the result 
\be
\Delta_{\rm pol} = 1.88 \pm 0.64 {\rm\ ppm} \,.
\ee
For other choices of $F_2(Q^2)$, the resulting changes in $\Delta_{\rm pol}$ are small compared to the overall error limit quoted above, as may be seen in the $\Delta_{\rm pol}$ column of Table~\ref{table:one}.  The current result differs from our previous $1.3\pm 0.3$ result~\cite{Nazaryan:2005zc} based on earlier data and a less sufficient treatment of the systematic errors.  Other $\Delta_{\rm pol}$ results incompatible with zero are the 2002 Faustov and Martynenko~\cite{Faustov:yp} value of $1.4\pm 0.6$ ppm and the 2006 Faustov, Gorbacheva, and Martynenko~\cite{Faustov:2006ve} value of $2.2 \pm 0.8$ ppm.


\subsection{Ordinary hydrogen Zemach and recoil corrections}


Form factor measurements have improved in the past decade largely due to the use of polarization transfer techniques~\cite{Gayou:2001qd} and to an understanding of how two-photon corrections impact the Rosenbluth measurements~\cite{Blunden:2003sp}.  Analytic form factor parameterizations new within the past year are available from Arrington, Melnitchouck, and Tjon~\cite{Arrington:2007ux}, who fit over all $Q^2$ where there is data, and from Arrington and Sick~\cite{Arrington:2006hm}, who concentrate on the lower $Q^2$ data.  The Zemach contributions from these, and two slightly older fits~\cite{Kelly:2004hm,Friedrich:2003iz}, are listed in the third and fourth columns of Table~\ref{table:one}.   One notices that modern form factors give larger radii and larger magnitude Zemach corrections than the old dipole form.

\begin{table}[b]

\caption{Zemach radii, $\Delta_Z$ including $\delta_Z^{\rm rad}$, and the recoil corrections, for four modern form factors.  The dipole form factor is included only as a benchmark.  The ``target'' $\Delta_S$ is $-32.77\pm0.01$ ppm;  the errors on $\Delta_S$ in the Table above are typically $\pm 0.7$ ppm.
}

\begin{ruledtabular}

\begin{tabular}{lcccccc}
Form factor  &$r_P$ &  $ r_Z$ & $\Delta_Z$	&$\Delta_R^p$ &$\Delta_{\rm pol}$ & $\Delta_S$\\
	& (fm)	& (fm)	& (ppm)	& (ppm)	& (ppm) & (ppm)	\\
AMT~\cite{Arrington:2007ux}  	& $0.885$	&  $1.080$ &  $-41.43$	& $5.85$	& $1.88$&$-33.70$	\\
AS~\cite{Arrington:2006hm}    	& $0.879$	&  $1.091$ &  $-41.85$ 	& $5.87$	& $1.89$&$-34.09$	\\
Kelly~\cite{Kelly:2004hm}		& $0.878$	&  $1.069$ &  $-40.99$	& $5.83$	& $1.89$&$-33.27$	\\
FW~\cite{Friedrich:2003iz}	& $0.808$	&  $1.049$ &  $-40.22$	& $5.86$	& $2.00$&$-32.36$	\\
dipole                        		& $0.851$	&  $1.025$ &  $-39.29$	& $5.78$	& $1.94$&$-31.60$	\\
\end{tabular}

\label{table:one}

\end{ruledtabular}

\end{table}


(The fits of Ref.~\cite{Arrington:2006hm} are valid only for $Q^2 \le 1$ GeV$^2$, and for $Q^2$ above this value we supplement them with form factors taken from Ref.~\cite{Arrington:2007ux}.  The integrals are strongly weighted to lower $Q^2$, so that if we supplement them with the dipole forms instead, the results would be the same to the number of figures given.)

Using the same form factors, we quote the recoil corrections in the fifth column of Table~\ref{table:one}.  The bulk of the result comes from the one-loop corrections of Eq.~(\ref{eq:recoil}).  We also included a $0.46$ ppm two-loop recoil correction from Bodwin and Yennie~\cite{Bodwin:1987mj}.  The latter are ${\mathcal O}(\alpha^2)$ beyond the Fermi energy scale and are given by,
\begin{eqnarray}
\label{eq:by2loop}
\Delta_{R}^p({\rm BY;\ }\alpha^2)&=& \alpha^2~ \frac{m_\ell}{m_p} \Bigg\{
   2 \ln { 1\over 2\alpha } - 6 \ln{2} + \frac{65}{18}
                    \nonumber \\
&&+~
 \kappa_p \left[
   \frac{7}{4} \ln { 1\over 2\alpha } - \ln{2} + \frac{31}{36}
 \right] \nonumber \\
&& \hskip -10 mm +~ {\kappa_p \over 1+\kappa_p} \left[
 -\frac{7}{4} \ln { 1\over 2\alpha } + 4 \ln{2} - \frac{31}{8}
 \right] \Bigg\}.
\end{eqnarray}
This correction is evaluated only in the static limit.  For the ${\mathcal O}(\alpha)$ recoil correction (leading order or LO), evaluating in the static limit gave a different sign and about a factor of $2$ smaller magnitude than using physical form factors.  However, this was possible only because the static evaluation of the LO is unexpectedly small due to striking internal cancellations.  To wit, the LO static correction evaluated with the measured $\kappa_p$ is about 15 times smaller than it would be using $\kappa_p = 0$.  Similar internal cancellations do not occur in the next-to-LO corrections, and we should not expect---albeit this is not verified---that evaluation with physical form factors would occasion big changes in the two-loop corrections.   We also include an additional $0.09$ ppm radiative correction noted by Karshenboim~\cite{Karshenboim:1996ew}.

An overall summary of calculated results for the ordinary hydrogen hfs is given in Table~\ref{table:deficit}, along with, in the first line, the experimental value of the corrections in units of the Fermi energy.   

Regarding error limits for the Zemach correction,  three of the four modern form factor fits give uncertainties in their fit parameters that allow an estimate of the uncertainty in $\Delta_Z$ obtained from the respective fits.  There are, of course, correlations.  For example, some of the data are cross sections, so that if the extracted $G_E$ goes up, then $G_M$ goes down.  We estimated the uncertainty in $\Delta_Z$ by letting $G_E$ vary to the maximum allowed by the respective authors's error limits, and doing so leads to variations in $\Delta_Z$ of $\pm 0.085$ ppm, $\pm 0.33$ ppm, and $\pm 0.80$ ppm for the AS~\cite{Arrington:2006hm}, Kelly~\cite{Kelly:2004hm}, and FW~\cite{Friedrich:2003iz} fits, respectively.  This may be an argument for favoring the AS fit.  However, the variations among the results for the different form factor fits are larger than some of the uncertainties just quoted, and we have taken the approach of using the result from the AMT fit~\cite{Arrington:2007ux} with an uncertainty chosen to accommodate the two most modern of the other fits.  A similar choice has been made for the recoil corrections.

The total of the hfs corrections gives a result that is $0.85$ ppm short of the data, with a quoted uncertainty of $0.78$ ppm.  The goal of a 1 ppm calculation appears to have been reached, with the theory versus data difference barely over a standard deviation.  There is no evidence of missing physics at this level.   Also, the uncertainty in the polarizability term is now comparable to the uncertainty in the Zemach term, which is purely  dependent upon the elastic form factors.

\begin{table}[h]
\caption{Summary of corrections for electronic hydrogen; $\Delta_Z$, $\Delta_R^p$, and $\Delta_{\rm pol}$ come from Tables~\ref{table:pol} and~\ref{table:one}.}

\begin{ruledtabular}
\begin{tabular}{lrc}
Quantity						&	value (ppm)	& uncertainty (ppm) \\
$({E_{\rm hfs}(e^-p)}/{E_F^p})  - 1$	&	$1\ 103.48 \ \ $  &	$0.01$	\\
\hline
$\Delta_{\rm QED}$				& 	$1\ 136.19 \ \ $	&	$0.00$	\\
$\Delta_{\mu{\rm vp}}^p+\Delta_{\rm hvp}^p
+\Delta_{\rm weak}^p$			&	$0.14 \ \ $		&	 		\\[1.3ex]
$\Delta_Z$ (using~\cite{Arrington:2007ux})	&	$-41.43 \ \ $	&	$0.44$	\\
$\Delta_R^p$ (using~\cite{Arrington:2007ux})	&	$5.85 \ \ $		&	$0.07$	\\
$\Delta_{\rm pol}$ (this work, using~\cite{Arrington:2007ux})		
									&	$1.88 \ \ $		&	$0.64$	\\
\hline
Total							&  $1102.63 \ \ $	&	$0.78$	\\
Deficit						&	$0.85 \ \ $		&	$0.78$     \\
\end{tabular}
\end{ruledtabular}
\label{table:deficit}
\end{table}


\subsection{Muonic hydrogen structure-dependent corrections}


For muonic hydrogen we, of course, keep $m_\ell \ne 0$. There are no poles in the integrands of Eqs.~(\ref{eq:DeltaMassive});  the numerators of the $\tilde\beta_i$ are zero when the denominators are.  For numerical purposes, one can analytically divide to obtain
\ba
\tilde\beta_1(\tau,\tau_\ell)	&=&
		-2\tau_\ell \tau + \frac{ \sqrt{\tau+1} }{ \sqrt{\tau_ \ell} + \sqrt{\tau} }
		\left( 2 \tau_ \ell \tau + 4 \sqrt{ \tau_ \ell \tau} \right)
						\nonumber	\\
		&+& \frac{ \sqrt{\tau_ \ell} }{ \sqrt{\tau_ \ell +1} + \sqrt{\tau+1} }
		\left( 2 \tau_ \ell \tau - 4 \tau  \right)	,
							\\[1.5ex]
\tilde\beta_2(\tau,\tau_\ell)
	&=& {2\tau_ \ell} 
						\nonumber \\
	&\times& \bigg\{ -1 + \frac{ \sqrt{\tau_ \ell +1} }{ \sqrt{\tau_ \ell} + \sqrt{\tau} }
		+  \frac{ \sqrt{\tau} }{ \sqrt{\tau_ \ell +1} + \sqrt{\tau+1} }
	\bigg\}		.			\nonumber
\ea

Evaluating $\Delta_i$ for the $0 < Q^2 < Q_1^2 = 0.0452$ GeV$^2$ data gap is somewhat different from the electron case.  Now, $\tau_\ell = \tau_\mu$ and is small (in the range 0 to about 1) rather than very large, although $\tau$ is still fairly large.  A numerically good approximation for these ranges is
\be
\tilde\beta_1(\tau,\tau_\mu)
	\approx \beta_1(\tau_\mu) \left( 1 - \frac{1}{6\tau} \right)  \,.
\ee
This leads to 
\ba
\Delta_1[0,Q_1^2] &=& \left( -\frac{1}{3} \kappa_p^2 r_P^2 + 8 m_p^2 c_1 - \frac{m_p^2}{3\alpha} \gamma_0  \right) 
					\nonumber \\
					&\times& \int_0^{Q_1^2} dQ^2 \beta_1(\tau_\mu) \,,
\ea
where $\gamma_0$ is the forward spin polarizability,
\be
\gamma_0 = \frac {2 \alpha} {m_p} \int_{\nu_{th}}^\infty  \frac{d\nu}{\nu^4} \ g_1(\nu,0)
\ee
and $\nu_{th} = m_\pi + (m_\pi^2 + Q^2)/(2 m_p)$.   From data, $\gamma_0 = [-1.01 \pm0.08({\rm stat}) \pm 0.10({\rm syst})] \times 10^{-4}$ fm$^4$~\cite{Drechsel:2002ar}, and $c_1 = 4.50 \pm 0.35 \pm 1.42$ is from the analog of Eq.~(\ref{eq:b1expansion}) but for $\Gamma_1(Q^2) = \int_0^{x_{\rm th}} dx \, g_1(x,Q^2)$~\cite{Nazaryan:2005zc}.

The evaluation of the polarizability corrections for higher $Q^2$ is similar to the evaluation in the electronic case, and depends upon the same combination described previously of EG1 data for $g_1/F_1$, Christy and NMC/E143 fits for $F_1$, and supplements from Simula {\it et al.} and E155 fits where there is no EG1 data~\cite{Dharmawardane:2006zd,Christy:2007ve,Arneodo:1995cq,Abe:1998ym,Simula:2001iy,Anthony:2000fn}.

Table~\ref{table:pol_muon} shows the breakdown of contributions to $\Delta_{\rm pol}$ for muonic hydrogen using the AMT~\cite{Arrington:2007ux} elastic form factor.   Error limits in the Table are combined the same way as for Table~\ref{table:pol}.  Finally combining the statistical, systematic, and model dependent errors in quadrature yields
\be
\Delta_{\rm pol} = 351 \pm 114 {\rm\ ppm}.
\ee
Results for $\Delta_{\rm pol}$ using other form factors are, as in the electron case, not greatly different on a scale set by the current systematic errors on $\Delta_{\rm pol}$.  Results are shown in Table~\ref{table:zemach_recoil_muon}.

\begin{table}[t]
\caption{Contributions to $\Delta_{\rm pol}$ for muonic hydrogen. As in Table~\ref{table:pol}, statistical, systematic, and modeling errors are given in the parentheses. }
\begin{ruledtabular}
\begin{tabular}{llcc}
Term		& $Q^2$ (GeV$^2$)	&  From  &  Value w/AMT~\cite{Arrington:2007ux} $F_2$	\\
\hline
$\Delta_1$	& [0, 0.0452]	&	$F_2$ and $g_1$	& $\ \  0.86 (0.17) (0.67) \ \ (\ )\ \  $	\\
			& [0.0452, 20]	&	$F_2$		& $\ \  6.77 \ \ (\ )\ \  (0.21) \ \ (\ )\ \  $	\\
			& 			&	$g_1$		& $\ \  0.18 (0.18) (1.62) (0.64) $	\\
			& [20, $\infty$]	&	$F_2$		& $\ \  0.00 \ \ (\ )\ \  (0.00) \ \ (\ )\ \  $	\\
			&			&	$g_1$		& $\ \  0.11 \ \ (\ )\ \ \ \ (\ )\ \  (0.01)$\\[1.2ex]
total $\Delta_1$&			&				& $\ \  7.92 (0.25) (2.30) (0.66) $ \\[1.5ex] 
$\Delta_2$	& [0, 0.0452]	&	$g_2$		& $ -0.12 \ \ (\ )\ \ \ \ (\ )\ \  (0.12)  $		\\
			& [0.0452, 20]	&	$g_2$		& $ -0.29 \ \ (\ )\ \ \ \ (\ )\ \  (0.29)  $		\\
			& [20, $\infty$]	&	$g_2$		& $ -0.00 \ \ (\ )\ \ \ \ (\ )\ \  (0.00)  $\\[1.2ex]
total $\Delta_2$&			&				& $ -0.41 \ \ (\ )\ \ \ \ (\ )\ \  (0.41)  $\\[1.5ex]
$\Delta_1 + \Delta_2$	&	&				& $\ \  7.51 (0.25) (2.30) (0.77) $	\\
$\Delta_{\rm pol}$ (ppm)		&	&			& $\ \  351. (\,12.\ ) (107.) (\,36.\ ) $
\end{tabular}
\end{ruledtabular}
\label{table:pol_muon}
\end{table}%

Also in Table~~\ref{table:zemach_recoil_muon} are results for the Zemach and recoil corrections in the muon case.   The structure dependent corrections become large compared to the electron case since they are, unlike the QED corrections, proportional to the lepton mass.  The Zemach corrections follow simply from scaling the electron case with the new reduced mass.  The recoil corrections are easily recalculated and include the two-loop corrections of Bodwin and Yennie quoted in Eq.~(\ref{eq:by2loop}).  The latter scale directly with the lepton mass; they were $0.46$ ppm for the electron case and are here $96$ ppm.   The extra radiative recoil corrections that accounted for $0.09$ ppm in the electron case have been omitted.  The vacuum polarization part of these corrections are easy to scale to the muon case~\cite{Karshenboim:1996ew}, but formulas are not available for the self energy part.  These corrections were, for electronic hydrogen, small compared to the current overall accuracy of the final result.

\begin{table}[b]

\begin{ruledtabular}

\caption{For muonic hydrogen hyperfine splitting: Zemach radii (as before, included for completeness), $\Delta_Z$ including $\delta_Z^{\rm rad}$, recoil corrections, polarizability corrections, and the summed structure dependent corrections $\Delta_S$, for four modern form factors, with the dipole form factor included as a benchmark.  Typical errors on $\Delta_S$ are $\pm 120$ ppm.
}

\begin{tabular}{lccccc}
Form factor   &  $ r_Z$ & $\Delta_Z$	&$\Delta_R^p$ &$\Delta_{\rm pol}$ & $\Delta_S$\\
		& (fm)	& (ppm)	& (ppm)	& (ppm) & (ppm)	\\
AMT~\cite{Arrington:2007ux}  &  $1.080$  &  $-7703.$	& $931.$ 	& $351.$	&$-6421.$		\\
AS~\cite{Arrington:2006hm}    &  $1.091$  &  $-7782.$	& $931.$	& $353.$	&$-6498.$		\\
Kelly~\cite{Kelly:2004hm}		&  $1.069$ & $-7622.$	& $931.$	& $353.$	&$-6338.$		\\
FW~\cite{Friedrich:2003iz}	&  $1.049$ & $-7479.$	& $939.$	& $370.$	&$-6170.$		\\
dipole                        		&  1.025	&  $-7311.$ 	& $935.$	& $362.$	&$-6014.$		\\
\end{tabular}

\label{table:zemach_recoil_muon}

\end{ruledtabular}

\end{table}


%
\subsection{Estimates regarding $\Delta_2$ }		\label{subsec:deltatwo}
%
%

In this subsection we reconsider $\Delta_2$, first using the Wandzura-Wilczek approximation~\cite{Wandzura:1977qf}, and then considering what existing proton data can tell us about the non-Wandzura-Wilczek part of $\Delta_2$.  After reconsideration, we shall still believe that $\Delta_2$ based on the EG1 model, with $100$\% error limits, is satisfactory for the present hfs accuracy goal.

The structure function $g_1$ can be divided into
\begin{equation}
g_2(x,Q^2) = g_2^{WW}(x,Q^2) + \bar{g}_2(x,Q^2)  \,,
\end{equation}
where the Wandzura-Wilczek relation states
\begin{equation}
g_2^{WW}(x,Q^2) = - g_1(x,Q^2) + \int_x^{x_{\rm th}} \frac{g_1(y,Q^2)}{y} dy  \,.
\end{equation}
Although at high $Q^2$, $g_2^{WW}$ is the leading twist contribution and
$\bar{g}_2$ is the higher twist component, this formal division is well-defined
all the way to $Q^2=0$.   Hence, we can define
\begin{eqnarray}
B_2(Q^2) &=& B_2^{WW} + \bar{B}_2	\,,
			\nonumber \\
B_2^{WW}(Q^2) &=& \int_0^{x_{\rm th}} dx \beta_2(\tau)g_2^{WW}(x,Q^2)		\,,
			\nonumber \\
\bar{B}_2(Q^2) &=& \int_0^{x_{\rm th}} dx \beta_2(\tau)\bar{g}_2(x,Q^2)	\,.
\end{eqnarray}
Substituting and manipulating yields
\begin{equation}
B_2^{WW} = \int_0^{x_{\rm th}} dx \beta_3(\tau) g_1(x,Q^2)	\,,
\end{equation}
in which
\begin{equation}
\beta_3(\tau) = 4\sqrt{\tau(\tau+1)} - 4\tau - 2\sqrt{\tau}
\ln\left(\frac {\sqrt{\tau+1}+1}{\sqrt{\tau}}\right)  ,
\end{equation}
for the electron case.
The function $\beta_3$ has limits
\begin{eqnarray}
\beta_3(\tau) &=& (\ln\tau - 2\ln 2 + 4)\sqrt{\tau} - 4\tau	\,,
		\quad \tau\to 0	,
				\nonumber	\\
\beta_3(\tau) &=& - \frac{1}{6\tau} + \frac{1}{10\tau^2} + ...  ,
		\qquad \tau\to\infty	\,.
\end{eqnarray}
The result is $\Delta_2^{WW}=-0.71\pm 0.08 \pm 0.10 \pm 0.01$ using the
same data and techniques as for determining $B_1$ and $\Delta_1$.

For the muon case, one replaces $\beta_3$ by
\begin{eqnarray}
\tilde\beta_3(\tau,\tau_\ell) &=& - \tilde\beta_2(\tau,\tau_\ell) - 2 \tau_\ell
	+ 2 \sqrt{\tau_\ell(\tau_\ell + 1)}  \nonumber \\[1ex]
&+& 2 \sqrt{\tau_\ell(\tau_\ell + 1)} \ln \left(\frac
	{\sqrt{\tau_\ell + 1}  + \sqrt{\tau + 1}} { \sqrt{\tau_\ell} + \sqrt{\tau}} \right)
								\nonumber \\
&-& 2\sqrt{\tau} \ln\left(\frac {\sqrt{\tau+1}+1}{\sqrt{\tau}}\right)   ,
\end{eqnarray}
and obtains $\Delta_2^{WW}\! (\mu p)=-0.57\pm 0.06 \pm 0.10 \pm 0.01$.
 
Notice that we obtain negative $\Delta_2$, which means that the main support in the integrals comes from regions where $g_2(x,Q^2)$ is positive.  Hence, the main contribution to the hfs from the $g_2$ terms comes from the resonance region, and specifically from the region of the $\Delta(1232)$ resonance, since the existing data shows $g_2$ is positive there and negative elsewhere.  As the integrals also have stronger support at low $Q^2$, one specific need for more data would be in the higher resonance and continuum regions at low $Q^2$.

There is interest among hadronic physicists in the higher twist component $\bar g_2(x,Q^2)$.  In particular, the higher twist coefficient defined by 
\begin{equation}
d_2(Q^2) \equiv 3\int_0^{x_{\rm th}} x^2 \, \bar{g}_2(x,Q^2) \, dx 
		= \int_0^{x_{\rm th}}	x^2  (2g_1 + 3g_2)dx
\end{equation}
has been studied by experimenters.

Osipenko {\it et al.}~\cite{Osipenko:2005nx} quote results for $d_2(Q^2)$ that are small at low and high $Q^2$, but significant and
positive for values within a decade on either side
of $Q^2=1$ GeV$^2$.  However, they also give systematic errors, and these are very large.
For comparison, Kao {\it et al.}~\cite{Kao:2003jd} also model $d_2$, and do not find a fixed sign.

At one value of $Q^2$, namely $1.3$ GeV$^2$, there is good data on the proton's $g_2(x,Q^2)$ from the RSS collaboration~\cite{Wesselmann:2006mw}.  This data shows that $g_2$ is typically about half of $g_2^{WW}$,  at this $Q^2$.  If this be generally true then $\Delta_2$ is, of course, about half the $\Delta_2^{WW}$ values just quoted.

We conclude that, partly because of the smallness of the contribution, the existing data allows us to use the $\Delta_2$ values quoted in our Tables, with confidence that the generous percentage error limits will include any changes that will come with better data.  Of course, we do want more complete data (there is already more complete data for neutron targets, extracted from polarized $^3$He targets) and know that experimenters are also interested.


\section{Conclusions}						\label{sec:end}


Our new result for the polarizability corrections to the hfs of the hydrogen ground state is,
\be
\Delta_{\rm pol} = 1.88 \pm 0.64 {\rm\ ppm} \,,
\ee
where the error limit includes both statistical and systematic uncertainties.  The main new ingredient in this determination is the recently published data on the spin-dependent structure function $g_1(\nu,Q^2)$ from the JLab EG1 collaboration~\cite{Dharmawardane:2006zd}.  The result is somewhat larger than our previously published $\Delta_{\rm pol}$~\cite{Nazaryan:2005zc}, and the quoted uncertainty limit is also larger due to a better comprehension of the systematic error.  

There are also recent results on $\Delta_{\rm pol}$ from Faustov, Gorbachova, and Martynenko~\cite{Faustov:2006ve}.  They have not used the EG1 data, relying instead on theoretically motivated fits to earlier data.  Their result is somewhat larger than ours, but compatible within error limits.

A consequence of the slightly larger $\Delta_{\rm pol}$ and larger uncertainty is that the calculated hfs is just within $1$ ppm, and just about one standard deviation, from the experimental value.  

The goal of a $1$ ppm hydrogen hfs calculation appears to be realized.  One needs to make this claim with some diffidence; a well-known~\cite{Bodwin:1987mj} paper is sometimes read as having made this claim in 1988.  However, the claim only referred to the accuracy of the methods, and the authors themselves pointed out that the polarizability correction at that time was known only as compatible with zero to the $4$ ppm level, and that the dipole form factor they used already differed systematically from the data.  Indeed, Ref.~\cite{Karshenboim:1996ew} found that a better low $Q^2$ fit to the form factor data changed the Zemach contributions by about $2$ ppm.  Now, both form factors and structure functions are better known, and a claim of $1$ ppm accuracy is plausible, with, at this level, no unknown terms remaining to be included.

Further improvement in the calculation of hfs using electron scattering data depends upon further improvement in the data and/or its analysis.  The largest uncertainty currently follows from systematic uncertainties in the inelastic structure functions.  In Tables~\ref{table:pol} and~\ref{table:pol_muon} we give error limits separately for the statistical, systematic, and modeling errors.  "Systematic" here means systematic errors only from the data, and "modeling" is the uncertainly estimated from the models that we use to complete the integrals where there is no data.  The statistical errors are small.  The largest errors are the systematic ones,  which can only be improved by understanding the apparatus better, or by improved apparatus.  

In addition, the uncertainty in the Zemach term, which depends upon elastic form factors, is also noticeable.  Even restricting to modern form factor fits, there is a $2$\% variation in the charge radius, with the fit~\cite{Arrington:2006hm,Sick:2003gm} arguably most attentive to the low $Q^2$ data giving the largest result.  Progress may depend not only on experimental progress, but also on a clearer understanding of the corrections needed to connect electron scattering cross section and polarization data to the form factors, and an assessment of how these corrections are implemented in current and future form factor fits.  Atomic determinations, based upon Lamb shift measurements, are in line with the lower values and have about $1$\% error limits~\cite{codata2006}.   Further, the atomic determinations of the charge radius will become remarkably more precise with the hoped for success of the muonic hydrogen Lamb shift experiment~\cite{Antognini:2005fe}.


The structure function $g_2(\nu,Q^2)$ is less important for the hfs than $g_1$, because the auxiliary function that multiplies it tends to be numerically small.  This is good, given that $g_2$ is harder to measure than $g_1$, and there is little data for the proton.  We included the $g_2$ contributions with $100$\% uncertainty limits and we believe these suffice.  The overall uncertainty still comes mainly from $g_1$.  We did give some further consideration to the $g_2$ contributions, in particular calculating the Wandzura-Wilczek part and discussing the remainder.  And certainly, more $g_2(x,Q^2)$ data would be welcome, to ensure that there are no surprises in, for example, the low $Q^2$ region at low $x$ (where the higher resonances and continuum contribute).

We have also given results for the structure dependent contributions to muonic hydrogen hfs. The total correction is unambiguous, but for the massive lepton case the protocols for separating the recoil and polarizability terms seem not yet standardized.  One hence needs to be watchful when adding together terms from different sources.  We have quoted algebraically all the structure dependent terms to leading correction order (order $\alpha \times$ mass ratio $\times$ Fermi energy) in the body of this paper, with matching conventions.  Our result for the polarizability term is
\be
\Delta_{\rm pol}(\mu^- p) = (351 \pm 114) {\rm\ ppm} \,,
\ee
and for the structure dependent terms overall, with the AMT elastic form factors,
\be
\Delta_{\rm S}(\mu^- p)  = \Delta_Z + \Delta_R^p + \Delta_{\rm pol} 
			= ( -6421 \pm 140 )  {\rm\ ppm} \,.
\ee
These are the biggest corrections for the muonic case.  The QED corrections, in particular, are very nearly the same size as in the electronic case, since they do not have the mass proportionality that the structure dependent terms have.

\begin{acknowledgments}
We thank Eric Christy for discussions and thank the National Science Foundation for support under grants PHY-0245056 (C.E.C.) and PHY-0653508 (V.N.), and the Department of Energy for support under contract DE-FG02-96ER41003 (K.A.G.).
\end{acknowledgments}

\end{document}